\documentstyle[11pt,paspconf,psfig,twoside]{article}

\begin{document}

\title{The optical counterpart of the double degenerate Polar RX J1914+24}

\author{Gavin Ramsay,$^{1}$ Mark Cropper,$^{1}$ Keith Mason,$^{1}$
Pasi Hakala$^{2}$}
\affil{$^{1}$Mullard Space Science Lab, University College London,
Holmbury St. Mary, Dorking, Surrey, RH5 6NT, UK\\
$^{2}$Observatory and Astrophysics Lab, FIN-00014, Univ Helsinki, Finland}

\begin{abstract}
Cropper et\,al.  presented X-ray observations of RX J1914+24 and
claimed that the most likely interpretation of their data was that of
a double degenerate Polar. Here we show the preliminary results of
optical and further X-ray observations. We identify the optical
counterpart and show that no other period apart from the 569 sec
period seen in X-rays is identified. Although the optical counterpart
has no significant polarised flux in the $I$ band, our {\sl ASCA}
spectrum (which is typical of a Polar) together with our photometric
result adds weight to the conclusion that RX J1914+24 is the first
double degenerate Polar to be discovered.
\end{abstract}


\section{Introduction}

Cropper et\,al. (1998) reported X-ray observations obtained using {\sl
ROSAT} of the X-ray source RX J1914+24 discovered by Motch et\,al.
(1996) and attribute the 569 sec period they detected to the binary
orbital period. This would make it the shortest orbital period of any
known binary system. From the absence of any other X-ray period and
the shape of the folded X-ray light curve, they suggest that RX
J1914+24 is the first double degenerate Polar to be discovered, rather
than the other possibility that it is an Intermediate Polar. This
paper gives the preliminary results of our search for the optical
counterpart of RX J1914+24 and our initial analysis of our {\sl ASCA}
X-ray spectrum of RX J1914+24.

\section{Photometry}

\begin{figure}
\begin{center}
\setlength{\unitlength}{1cm}
\begin{picture}(6,5)
\put(12,18.){\includegraphics{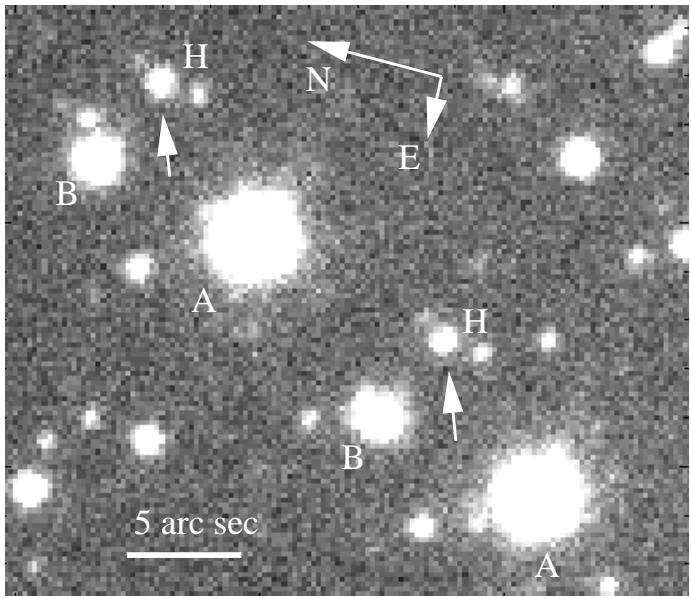}}
\end{picture}
\end{center}
\caption{An I band image of the field of RX 1914 taken in June 1998
using NOT. Two images of the each object in the field are recorded -
one from each polarised ray. Stars A and B in the finding chart of
Motch et\,al. (1996) are shown. RX J1914+24 is marked by
an arrow. Star H of Motch et\,al. is thus a blend of more than 1 star 
(including RX J1914+24).}
\label{noti} 
\end{figure}

Cropper et\,al. (1998) found that RX J1914+24 was closest to star H in
the finding chart of Motch et\,al. (1996). Motch et\,al. indicate that
star H is heavily reddened and not visible in $V$. We therefore
obtained a service image of RX J1914+24 in $K$ taken using UKIRT in
Oct 1997 and found that star H is made up of more than 1 star. Follow-up 
observations were obtained between 25--27 June 1998 using the NOT
on La Palma. Polarimetry observations were made in the $I$ band: o and
e rays were separated by $\sim12^{''}$ on the same CCD (fig
\ref{noti}).

Photometry of stars in the field was obtained using profile fitting.
As the profiles of the stars in the two rays were different,
photometry was obtained on each set of images. Our NOT photometry
showed that the brightest of the stars making up star `H' was the
optical counterpart of RX J1914+24. The mean magnitude was
$I\sim$18.2: $\sim$1.6 mag fainter than found by Motch et\,al. (1996) in
June 1993.  There is no evidence that RX J1914+24 is significantly
circularly polarised. In addition to these $I$ band data, $J$ band
data taken using UKIRT the following week have still to be analysed.

\section{The folded data}

\begin{figure}
\begin{center}
\setlength{\unitlength}{1cm}
\begin{picture}(8,8)
\put(-2,-2.){\includegraphics{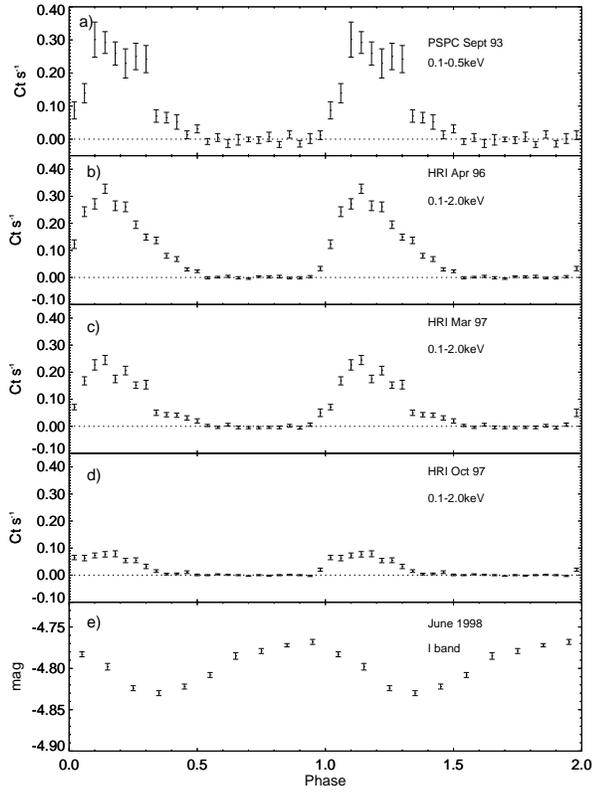}}
\end{picture}
\end{center}
\caption{The {\sl ROSAT} X-ray data from different epochs together
with the I band photometry taken in June 1998 folded on the ephemeris
T$_{o}$= HJD 2449258.03941 + 0.0065902334.}
\label{fold} 
\end{figure}

Cropper et\,al. (1998) showed {\sl ROSAT} data from 3 distinct
epochs. Since then we have obtained another set of {\sl ROSAT}
data. We have been able to derive an accurate ephemeris using the
start of the increase in X-ray flux to define $\phi$=0.0 (T$_{o}$= HJD
2449258.03941(6) + 0.0065902334(4)). Fig \ref{fold} shows all our
{\sl ROSAT} data folded on this ephemeris. This shows that there has
been a gradual decrease in X-ray flux from Sept 1993 to Oct 1997
(fig \ref{fold}). In other respects the light curves are similar in
that they all show zero flux for $\sim$ half the 569 sec period.

\begin{figure}
\begin{center}
\setlength{\unitlength}{1cm}
\begin{picture}(8,5)
\put(-4,-32.){\includegraphics{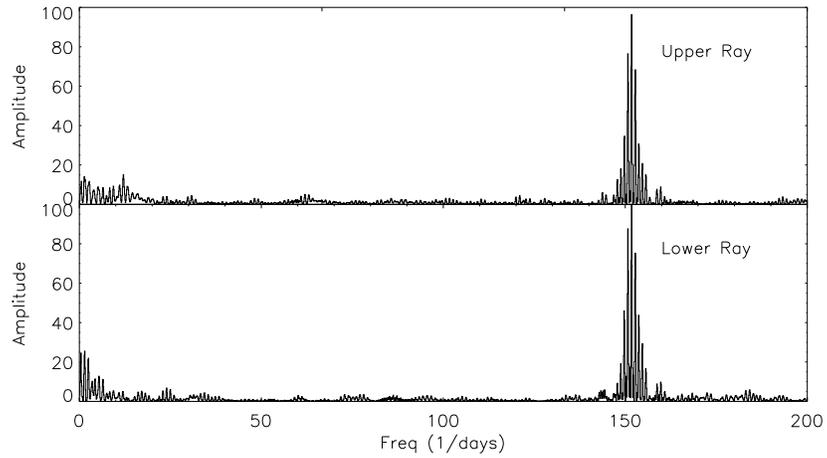}}
\end{picture}
\end{center}
\caption{The amplitude spectra of RX J1914+24 covering 3 nights of $I$
band photometry obtained using NOT in June 1998. Since both polarised
rays were recorded we obtained two amplitude spectra. The prominent
peak near Freq=150 is the 569 sec period seen in X-rays.}
\label{power} 
\end{figure}

A power spectrum of our $I$ band photometry indicates that only one
period (the 569 sec period seen in X-rays) is present (fig \ref{power}).
We show the $I$ band data folded using all 3 nights of data in the
bottom panel of fig \ref{fold} using the above ephemeris (which is
sufficiently accurate to match the phasing to within 0.02 cycles). The
folded data is roughly sinusoidal in shape. Interestingly, the peak
precedes the X-ray peak by $\phi\sim$0.4.

\section{The {\sl ASCA} spectrum}

\begin{figure}
\begin{center}
\setlength{\unitlength}{1cm}
\begin{picture}(8,7)
\put(-5,-6.){\includegraphics{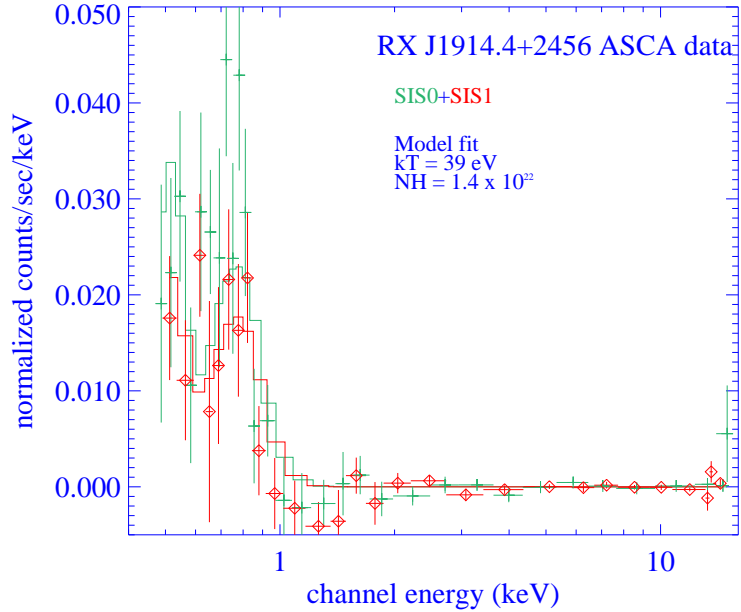}}
\end{picture}
\end{center}
\caption{The integrated {\sl ASCA} SIS-0 and SIS-1 spectra. The fit to
an absorbed blackbody model is shown together with their fitted parameters.}
\label{asca} 
\end{figure}

We obtained a 20ksec observation of RX J1914+24 in April 1998. RX J1914+24 was
detected only in the SIS and only below 1keV. Its spectrum (fig \ref{asca})
could be well fitted with a low temperature ($kT\sim$40eV) blackbody plus
interstellar absorption (N$_{H}\sim1\times10^{22}$ cm$^{-2}$). The $4\sigma$ 
upper limit
to a thermal bremsstrahlung component (assumed temperature 20keV)
corresponds to a $2-10$ keV flux of $3.2\times10^{-13}$ erg/cm$^{2}$/s
($4\times10^{29}$ erg/s at 100 pc). This is typical of a Polar. X-ray spectra
of Intermediate Polars are typically much harder.

\section{Conclusion}

We have detected the optical counterpart of RX J1914+24. We find only
one period in its $I$ band folded optical light curve (the same 569
sec period we find in X-rays). When we folded this data on the 569 sec
period we obtain a quasi-sinusoidal modulation offset in phase
compared with the X-rays. The fact that we detect only one period,
together with the very soft {\sl ASCA} spectrum is further evidence in
favour of the interpretation that RX J1914+24 is a double degenerate
Polar. A detailed paper is in preparation.


\begin{references}
\reference Cropper, M., et\,al., 1998, MNRAS, 293, L57
\reference Motch, C., et\,al., 1996, A\&A, 307, 459
\end{references}
\end{document}